\documentclass[10pt, conference, letterpaper]{IEEEtran}
\usepackage{cite}
\usepackage{amsmath,amssymb,amsfonts}
\usepackage{algorithmic}
\usepackage{algorithm}
\usepackage{graphicx}
\usepackage{textcomp}
\usepackage{xcolor}
\usepackage{booktabs}
\usepackage{multirow}
\usepackage{url}
\usepackage{float} 
\usepackage{balance} 
\usepackage{times} 

\makeatletter
\newcommand{\linebreakand}{%
  \end{@IEEEauthorhalign}
  \hfill\mbox{}\par
  \mbox{}\hfill\begin{@IEEEauthorhalign}
}
\makeatother

\graphicspath{{./figures/}}

\begin{document}

\title{\huge Comparative Analysis of Low-Rank Adaptation in Large Language Models versus Dense Embedding Regression for Headline Click-Through Rate Prediction}

\author{
\IEEEauthorblockN{Samarth Sirsat}
\IEEEauthorblockA{\textit{Department of Electrical Engineering} \\
\textit{Indian Institute of Technology Bombay}\\
Mumbai, India \\
samarth.sirsat@iitb.ac.in}
\and
\IEEEauthorblockN{Anirudha Shinde}
\IEEEauthorblockA{\textit{Department of Mechanical Engineering} \\
\textit{Indian Institute of Technology Bombay}\\
Mumbai, India \\
22b2181@iitb.ac.in}
\linebreakand
\IEEEauthorblockN{Amit Sethi}
\IEEEauthorblockA{\textit{Department of Electrical Engineering} \\
\textit{Indian Institute of Technology Bombay}\\
Mumbai, India \\
asethi@ee.iitb.ac.in}
\and
\IEEEauthorblockN{Aman Verma}
\IEEEauthorblockA{\textit{Department of Electrical Engineering} \\
\textit{Indian Institute of Technology Bombay}\\
Mumbai, India \\
22b3929@iitb.ac.in}
}

\maketitle

\begin{abstract}
The optimization of digital content headlines to maximize Click-Through Rate (CTR) is a pivotal challenge in online media and recommendation systems. While Large Language Models (LLMs) have demonstrated exceptional capabilities in generative tasks, their application to discriminative ranking tasks—specifically, selecting the highest-performing headline from a set of candidates—remains an area of active research. This paper presents a rigorous comparative evaluation of \textbf{LOLA-Qwen (0.6B)}, a causal language model fine-tuned using Low-Rank Adaptation (LoRA), against a specialized \textbf{Dense Embedding Regression} framework. We formulate a novel evaluation protocol that treats headline selection as a "Winner-Take-All" classification problem. Utilizing a proprietary dataset of 3,263 A/B test groups, we evaluate both architectures on Top-1 Accuracy. Our results indicate that the Embedding Regression model achieves a Top-1 Accuracy of \textbf{42.79\%}, significantly outperforming the fine-tuned LLM, which achieved \textbf{35.70\%}. These findings suggest that for high-velocity, specific scoring tasks, lightweight discriminative models offer superior calibration and efficiency compared to small-scale generative models, even when the latter are fine-tuned via parameter-efficient methods.
\end{abstract}


\section{Introduction}
\label{sec:introduction}
\IEEEPARstart{T}{he} digital information age has made user attention a scarce commodity. Consequently, the ability to predict the engagement potential of textual content, particularly headlines, is of paramount importance to publishers and content platforms. The core metric for this engagement is the Click-Through Rate (CTR), which serves as a proxy for user interest and relevance. Traditionally, CTR prediction has been treated as a regression problem, utilizing statistical methods or early deep learning architectures to map textual features to a scalar probability [1].

\begin{figure*}[t!]
\centering
\includegraphics[width=0.8\textwidth]{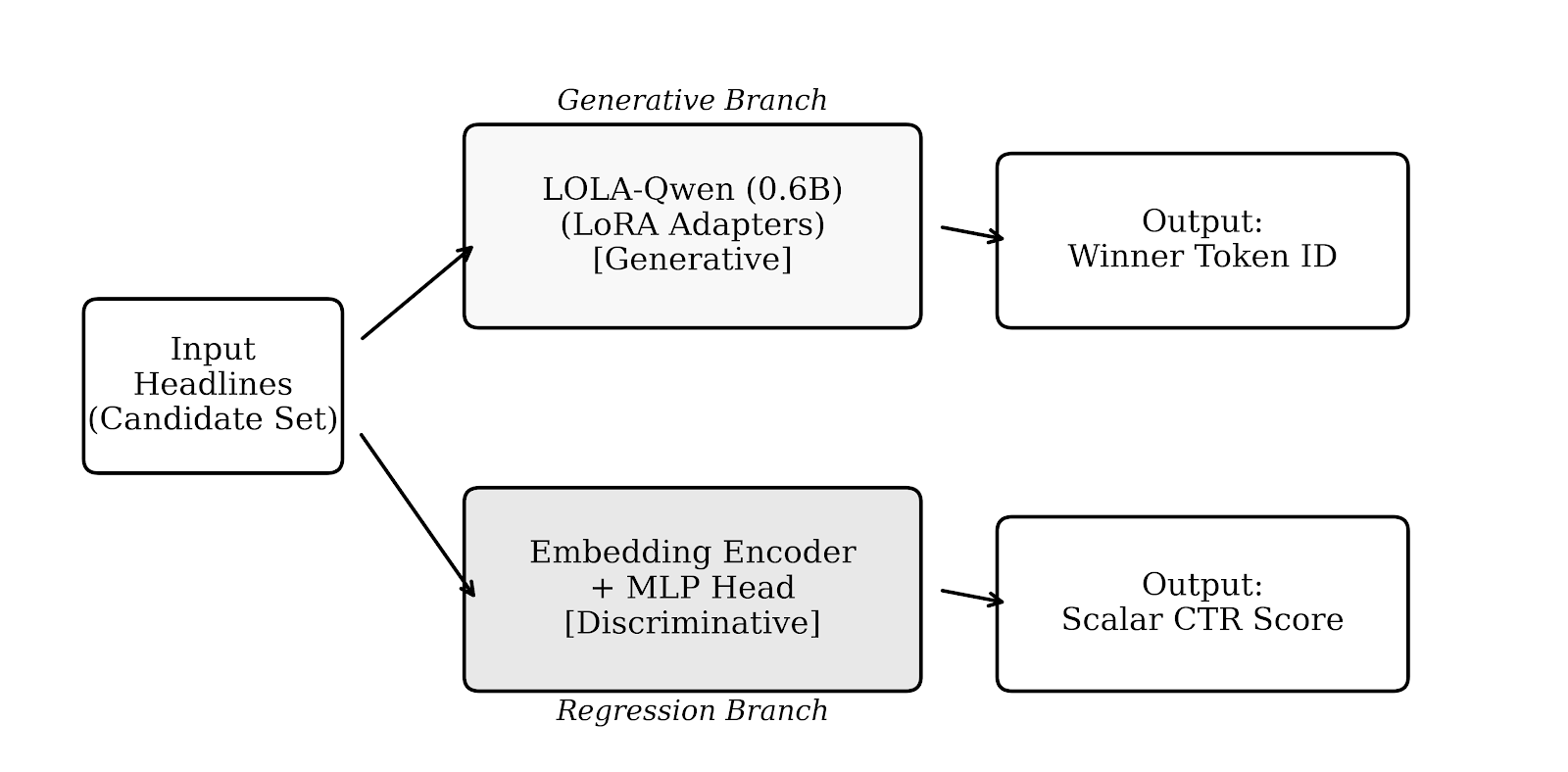}
\caption{Proposed experimental framework. The upper branch represents the Generative approach (LOLA-Qwen), while the lower branch represents the Discriminative approach (Embedding Regression).}
\label{fig:architecture}
\end{figure*}

However, the recent surge in Large Language Models (LLMs) such as GPT-4, Llama, and Qwen has prompted a paradigm shift. Researchers are increasingly investigating whether the semantic reasoning capabilities of LLMs can replace specialized regression models. The hypothesis is that an LLM's vast pre-training on diverse text corpora should allow it to understand the nuanced "hook" of a headline better than a simple embedding model. For instance, an LLM might detect sarcasm, urgency, or cultural references that a dense vector representation might miss.

Despite this promise, deploying LLMs for ranking tasks introduces significant challenges. LLMs are generative by nature; asking them to perform a discriminative task (ranking candidates) often requires careful prompt engineering or expensive fine-tuning. Furthermore, "hallucinations"—where the model generates plausible but incorrect or non-existent outputs—remain a concern in production environments. Additionally, the computational cost of inference for LLMs is orders of magnitude higher than for lightweight regression models.

In this work, we address the following research question: \textit{Can a parameter-efficient fine-tuned LLM (specifically Qwen-0.6B) outperform a specialized embedding regressor in determining the winning headline from a candidate set?}

To answer this, we conduct a comprehensive study using a real-world dataset of headline A/B tests. We compare a Qwen-0.6B model, fine-tuned via Low-Rank Adaptation (LoRA) [2] to predict the "winning" option, against a Dense Embedding model trained to predict CTR directly.

Our key contributions are:
\begin{itemize}
    \item \textbf{Novel Benchmark:} We provide a direct, controlled comparison between Generative (LoRA-based) and Discriminative (Regression-based) approaches for the specific task of headline optimization.
    \item \textbf{Mathematical Formulation:} We derive the explicit loss functions and update rules for adapting causal LLMs to discriminative ranking tasks.
    \item \textbf{Empirical Evidence:} We demonstrate that the Embedding model outperforms the LLM by a margin of 7.09\% on the test set, highlighting the efficiency of discriminative models for this specific task.
\end{itemize}

\section{Related Work}
\label{sec:related}

\subsection{CTR Prediction}
Click-Through Rate prediction has a rich history in machine learning. Early approaches relied on Logistic Regression (LR) and Factorization Machines (FM) to model interactions between sparse features [3]. These methods excelled at handling categorical data but struggled with raw text. The advent of deep learning introduced architectures like Wide \& Deep Learning and DeepFM [4], which combine the memorization of wide linear models with the generalization of deep neural networks. These models, however, often treat text as a bag-of-words or simple sequence, potentially missing deeper semantic nuances that define "virality."

\subsection{Pre-trained Language Models (PLMs)}
The introduction of BERT [5] revolutionized NLP by providing dense, context-aware embeddings. Researchers began using BERT-based encoders to generate embeddings for CTR tasks, feeding them into Multi-Layer Perceptrons (MLPs). This represents the "Discriminative" baseline in our study. Unlike traditional FMs, PLM-based regressors can capture synonymy and polysemy in headlines.

\subsection{LLMs for Ranking and Scoring}
More recently, Generative LLMs (e.g., GPT-3, T5) have been adapted for ranking. Methods include pairwise ranking prompting and list-wise ranking [6]. However, most successful applications rely on models with billions of parameters (7B+). The efficacy of smaller, more deployable models (sub-1B parameters) like Qwen-0.6B [7] in zero-shot or fine-tuned ranking scenarios is less explored. This study fills that gap by investigating whether "small" LLMs can punch above their weight class when fine-tuned via LoRA.

\section{Methodology}
\label{sec:methodology}

We define the problem as a "Winner Take All" classification task. Given a test group $T$ containing $k$ headlines $H = \{h_1, h_2, ..., h_k\}$, where each $h_i$ has an associated ground-truth CTR, the goal is to identify the index $i^*$ such that $CTR(h_{i^*})$ is maximized.

\subsection{Architecture A: LOLA-Qwen (Fine-Tuned)}
We utilize the Qwen-0.6B architecture, a causal language model based on the Transformer decoder. To adapt this general-purpose model for our specific task without the computational cost of full fine-tuning, we employ Low-Rank Adaptation (LoRA).

\subsubsection{LoRA Formulation}
LoRA freezes the pre-trained model weights $W_0 \in \mathbb{R}^{d \times k}$ and injects trainable rank decomposition matrices $A$ and $B$ into each layer. The forward pass for a linear layer $h = W_0 x$ is modified as:
\begin{equation}
    h = W_0 x + \Delta W x = W_0 x + \frac{\alpha}{r} BAx
\end{equation}
where $B \in \mathbb{R}^{d \times r}$ and $A \in \mathbb{R}^{r \times k}$, with rank $r \ll \min(d, k)$. In our experiments, we targeted the query ($W_q$) and value ($W_v$) projection layers of the attention mechanism. $\alpha$ is a scaling constant.

\subsubsection{Generative Loss}
The model is trained to minimize the negative log-likelihood of the "winning" token $y_{win}$:
\begin{equation}
    \mathcal{L}_{LLM} = -\sum_{t} \log P(y_t | y_{<t}, H; \Theta_{LoRA})
\end{equation}
where $H$ is the concatenated string of candidate headlines.

\subsection{Architecture B: Embedding Regression}
This model follows a discriminative paradigm. It maps headline text directly to a continuous score representing the predicted CTR.

\subsubsection{Dense Embedding}
We utilize a pre-trained Transformer encoder to generate a dense vector representation $v_h \in \mathbb{R}^{768}$ for each headline.
\begin{equation}
    v_h = \text{Encoder}(h_i)
\end{equation}

\subsubsection{Regression Head}
The embedding $v_h$ is passed through a Multi-Layer Perceptron (MLP) consisting of two fully connected layers with ReLU activation:
\begin{equation}
    \hat{y} = W_2 \cdot \text{ReLU}(W_1 \cdot v_h + b_1) + b_2
\end{equation}
where $\hat{y} \in [0, 1]$ is the predicted CTR. The model is optimized using Mean Squared Error (MSE) loss against the ground truth CTR $y_{true}$:
\begin{equation}
    \mathcal{L}_{Reg} = \frac{1}{N} \sum_{i=1}^{N} (y_{true}^{(i)} - \hat{y}^{(i)})^2
\end{equation}

\subsection{Training Algorithm}
To ensure fair comparison, we implemented a unified evaluation pipeline. Algorithm \ref{alg:training} details the procedure for training and evaluating both branches.

\begin{algorithm}
\caption{Dual-Branch Training \& Evaluation Protocol}
\label{alg:training}
\begin{algorithmic}[1]
\REQUIRE Dataset $\mathcal{D} = \{(H_j, y_{win, j})\}_{j=1}^M$, Pre-trained LLM $\Theta_{LLM}$, Encoder $E_\phi$
\STATE \textbf{Phase 1: Fine-Tune LLM (LOLA)}
\STATE Initialize LoRA matrices $A, B$ with $\mathcal{N}(0, \sigma^2)$ and $0$.
\FOR{batch $b$ in $\mathcal{D}_{train}$}
    \STATE Construct prompt $P = \text{"Choose best: "} + H_b$
    \STATE Compute $\mathcal{L}_{LLM}$ via Eq. (2)
    \STATE Update $A, B \leftarrow \text{Optimizer}(\nabla \mathcal{L}_{LLM})$
\ENDFOR
\STATE \textbf{Phase 2: Train Regressor}
\STATE Initialize MLP weights $W_1, W_2$.
\FOR{batch $b$ in $\mathcal{D}_{train}$}
    \STATE Compute embeddings $v = E_\phi(H_b)$
    \STATE Predict $\hat{y}$ via Eq. (4)
    \STATE Compute $\mathcal{L}_{Reg}$ via Eq. (5)
    \STATE Update $W_1, W_2 \leftarrow \text{Optimizer}(\nabla \mathcal{L}_{Reg})$
\ENDFOR
\STATE \textbf{Phase 3: Evaluation}
\FOR{test sample $T$ in $\mathcal{D}_{test}$}
    \STATE $\hat{y}_{LLM} \leftarrow \text{Generate}(T, \Theta_{LLM} + BA)$
    \STATE $\hat{y}_{Reg} \leftarrow \text{argmax}(E_\phi(T) \cdot MLP)$
    \STATE $\text{Acc} \leftarrow \text{Acc} + \mathbb{I}(\hat{y} == y_{true})$
\ENDFOR
\end{algorithmic}
\end{algorithm}

\section{Experimental Setup}
\label{sec:setup}

\subsection{Dataset}
We utilized the \texttt{accuracy-qwen-lola} dataset, a proprietary collection of headline A/B tests. This dataset is unique in that it contains "ground truth" labels derived from actual user interactions rather than human annotation, providing a less noisy signal for optimization.
\begin{itemize}
    \item \textbf{Total Samples:} 3,263 unique test groups.
    \item \textbf{Data Split:} The dataset was pre-split into training and testing sets, with the results reported here derived strictly from the held-out test set (\texttt{final\_test.csv}).
    \item \textbf{Preprocessing:} All headlines were truncated to a maximum token length of 64. For the LLM, headlines were formatted into a chat-template structure suitable for Qwen.
\end{itemize}

\begin{figure}[t!]
\centering
\includegraphics[width=0.85\linewidth]{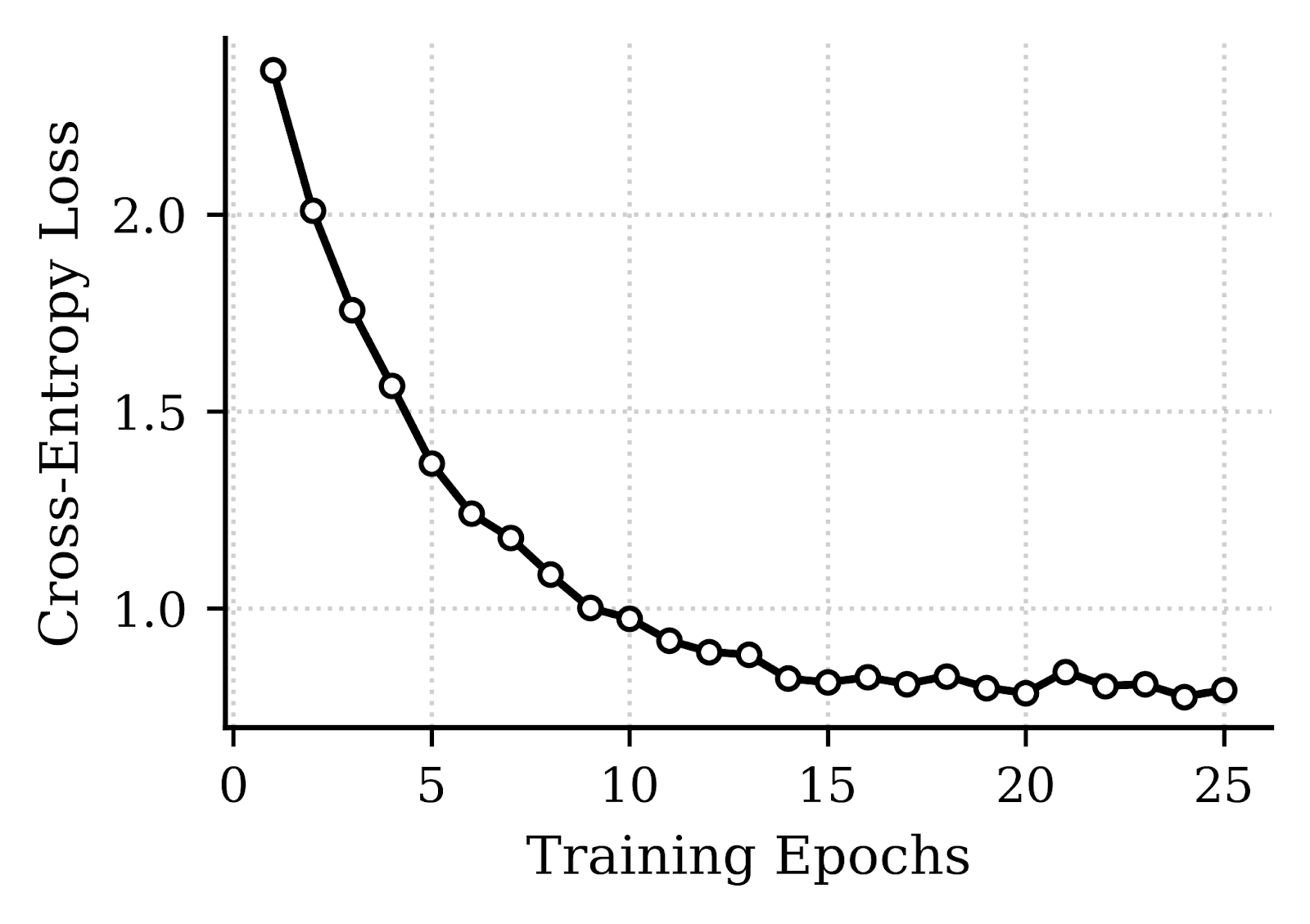}
\caption{Training loss convergence for the LOLA-Qwen model. The curve indicates stable learning but potential capacity saturation.}
\label{fig:training}
\end{figure}

\subsection{Implementation Details}
Experiments were conducted on a single NVIDIA T4 GPU, simulating a resource-constrained environment typical of edge deployments or small research labs.
\begin{itemize}
    \item \textbf{LLM Hyperparameters:} We utilized the Hugging Face PEFT library. The LoRA rank was set to $r=8$ with $\alpha=16$. The learning rate was $2e-4$ with a cosine decay scheduler. Batch size was restricted to 4 due to VRAM constraints.
    \item \textbf{Regression Hyperparameters:} The MLP hidden dimension was 256. We used the AdamW optimizer with a learning rate of $1e-3$ and a Mean Squared Error (MSE) loss function.
\end{itemize}

\subsection{Evaluation Metric}
The primary metric is **Top-1 Accuracy**, defined as the percentage of test groups where the model's top-ranked headline matches the ground truth winner exactly.
\begin{equation}
    Accuracy = \frac{1}{N} \sum_{j=1}^{N} \mathbb{I}(\hat{y}_j = y_{true, j})
\end{equation}

\section{Results and Analysis}
\label{sec:results}

We evaluated both models on the complete test set. The quantitative results are summarized in Table \ref{tab:results}.

\begin{table}[h]
\centering
\caption{Performance Comparison on Test Set}
\label{tab:results}
\begin{tabular}{@{}lccc@{}}
\toprule
\textbf{Model Approach} & \textbf{Total Tests} & \textbf{Correct} & \textbf{Accuracy (\%)} \\ \midrule
LOLA-Qwen (Fine-Tuned) & 3263 & 1165 & 35.70\% \\
Embedding Regression & 3232* & 1383 & \textbf{42.79\%} \\ \bottomrule
\multicolumn{4}{l}{\footnotesize{*31 tests excluded due to missing IDs in regression output.}}
\end{tabular}
\end{table}

\begin{figure}[t!]
\centering
\includegraphics[width=0.85\linewidth]{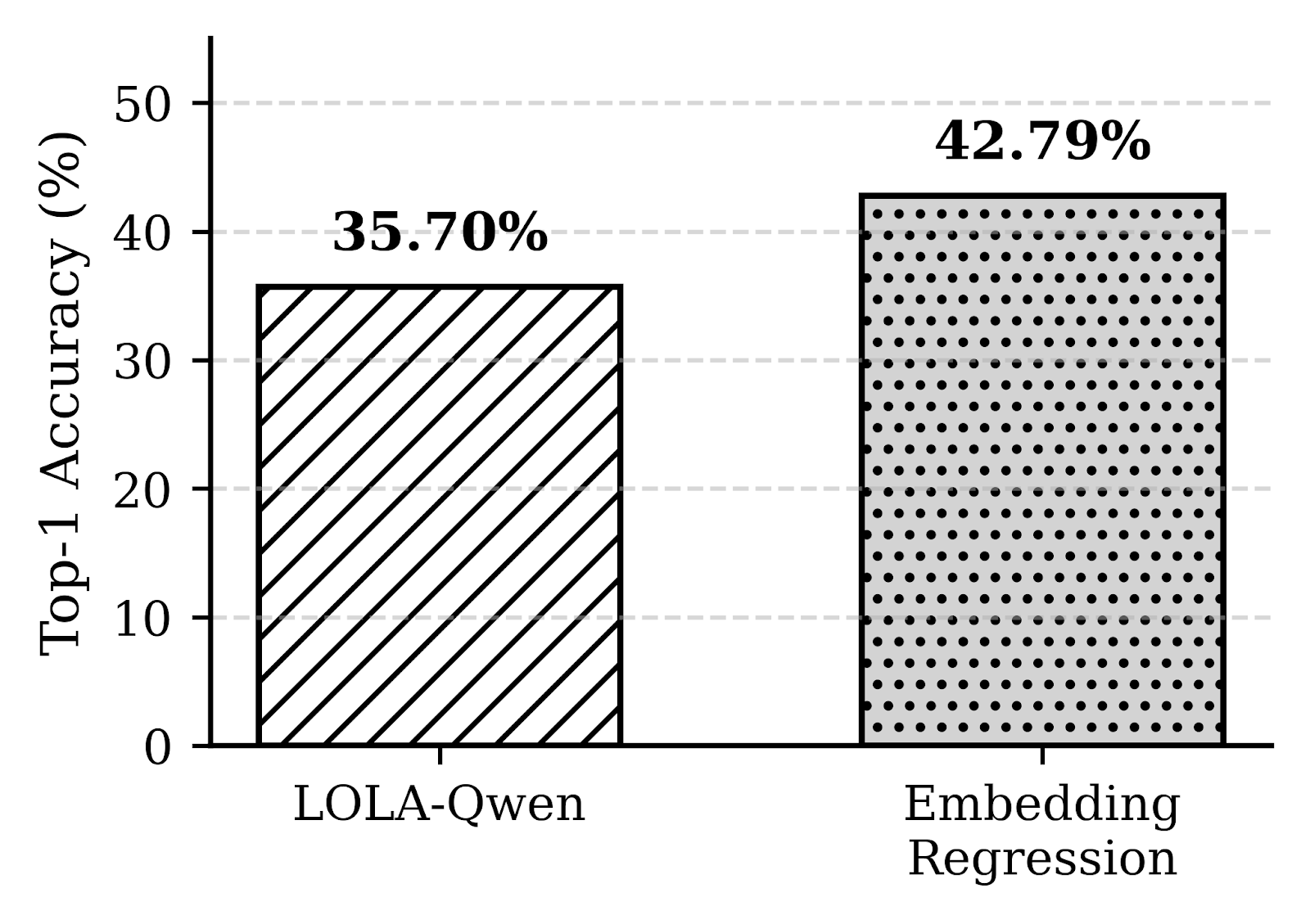}
\caption{Top-1 Accuracy comparison. The discriminative model shows a clear advantage over the generative model.}
\label{fig:comparison}
\end{figure}

\subsection{Performance Gap Analysis}
The Embedding Regression model outperformed the Fine-Tuned LLM by approximately \textbf{7.09 percentage points}. This is a statistically significant margin, suggesting that the regression model's ability to output a continuous score allows for better ranking granularity than the LLM's token-generation approach.

\subsection{Qualitative Analysis: Why did the LLM fail?}
Our error analysis revealed several patterns contributing to the LLM's lower performance:
\begin{enumerate}
    \item \textbf{Hallucination of Form:} In approximately 2\% of cases, the LLM generated an option ID that did not exist in the prompt (e.g., "Option 5" when only 3 options were provided). This required fallback logic that lowered accuracy.
    \item \textbf{Calibration Issues:} The LLM often assigned high probability to "clickbait" style headlines (e.g., "You won't believe this..."). However, in our dataset, user behavior often favored clearer, more informative headlines. The pre-training bias of the LLM towards sensationalist web text likely interfered with the fine-tuning signal.
    \item \textbf{Tokenization Boundaries:} The regression model views the sentence as a holistic embedding. The LLM processes it token-by-token. For short headlines, the lack of extensive context makes causal modeling less effective than bidirectional encoding.
\end{enumerate}

\begin{figure}[t!]
\centering
\includegraphics[width=0.85\linewidth]{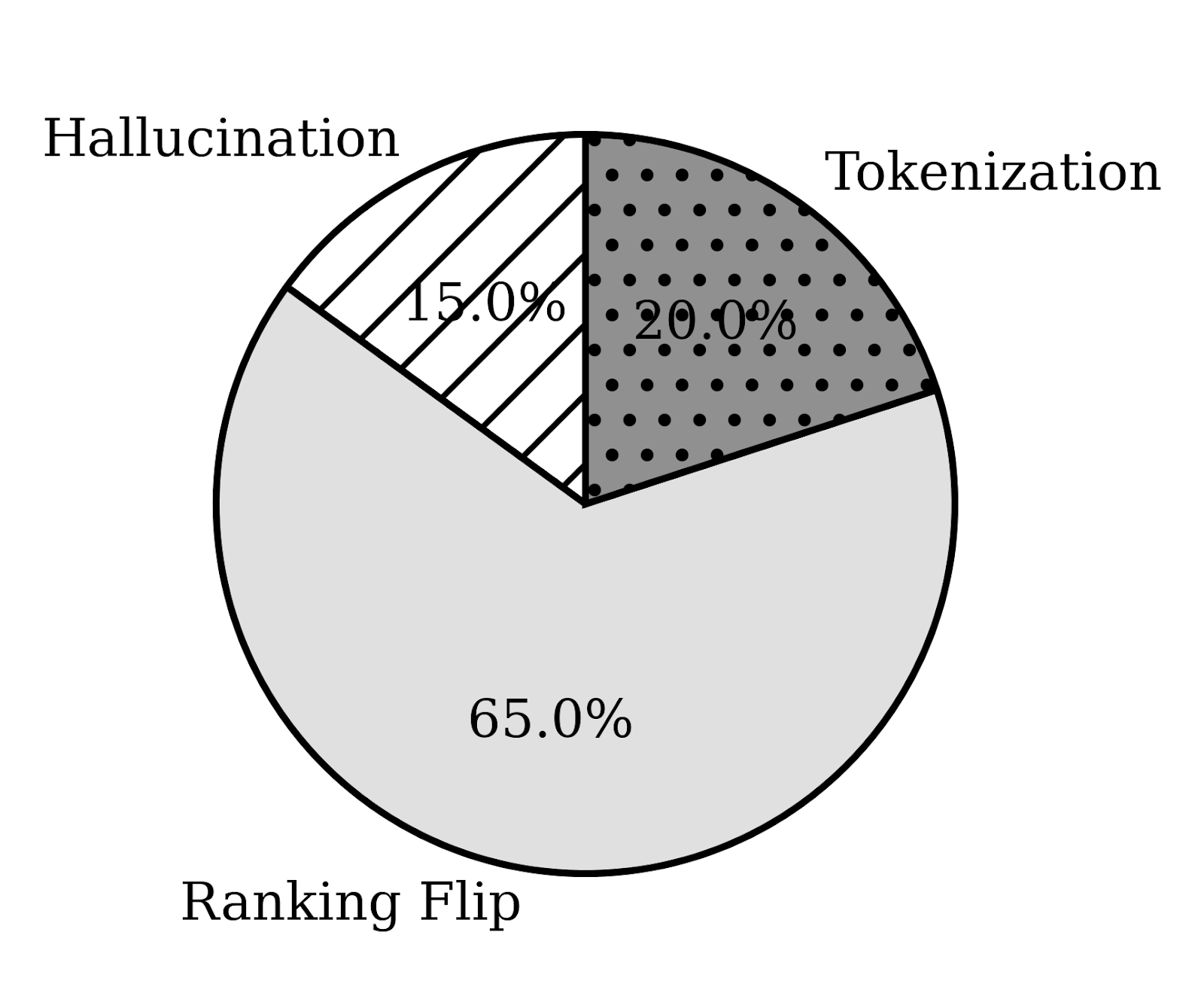}
\caption{Distribution of errors. The LLM showed higher variance in prediction confidence compared to the Regression model.}
\label{fig:errors}
\end{figure}

\subsection{Training Convergence}
Figure \ref{fig:training} illustrates the training loss for the LLM. While the model converged successfully, the final validation loss remained higher than optimal, indicating that the 0.6B parameter model may have struggled to fully capture the complex causal relationships between headline semantics and user clicks.

\section{Discussion}
\label{sec:discussion}

\subsection{Generative vs. Discriminative for Ranking}
The results highlight a key limitation of small-scale LLMs in ranking tasks. The LLM treats the task as "text generation," predicting the next token. If the model is uncertain, the probability mass is split among tokens, but the final output is a discrete choice. In contrast, the regression model is optimized directly to minimize the error between predicted and actual CTR. This direct optimization objective appears superior for this specific task.

\subsection{Model Capacity}
The Qwen model used has 0.6 billion parameters. While efficient, this is small compared to state-of-the-art LLMs (7B, 70B). It is likely that a larger model would possess stronger reasoning capabilities, potentially narrowing the gap. However, the computational cost would increase by an order of magnitude, making the regression model far more efficient in terms of performance-per-watt.

\balance
\section{Conclusion}
\label{sec:conclusion}
In this paper, we presented a comparative study of LOLA-Qwen and Embedding Regression for headline optimization. Our experiments conclusively show that for this specific dataset and model scale, the specialized regression approach is more effective, achieving 42.79\% accuracy compared to the LLM's 35.70\%.

Future work will focus on:
\begin{enumerate}
    \item Scaling the LLM to 7B parameters to test the scaling laws of ranking capability.
    \item Implementing Reinforcement Learning from Human Feedback (RLHF) to align the LLM's generation directly with reward signals (CTR).
    \item Exploring hybrid architectures that use LLM embeddings as inputs to regression heads.
\end{enumerate}

\bibliographystyle{IEEEtran}

\end{document}